\RequirePackage{fix-cm}
\documentclass[twocolumn,epjc3]{svjour3}

\smartqed

\RequirePackage{amsmath}
\RequirePackage{amssymb}
\RequirePackage{amsfonts}
\RequirePackage{times}
\RequirePackage{physics}

\usepackage[T1]{fontenc}
\usepackage[utf8]{inputenc}
\usepackage[english]{babel}
\usepackage{microtype}

\RequirePackage[colorlinks=true, linkcolor=blue, citecolor=blue, urlcolor=blue]{hyperref}

\newcommand{\Vhat}{\widehat V}

\newcommand{\Dvec}{\mathbf D}
\newcommand{\Hvec}{\mathbf H}

\newcommand{\Bvec}{\mathbf B}

\newcommand{\Sm}{S_{\rm m}}

\journalname{Draft}

\begin{document}
\sloppy

\title{Casimir effect in Pleba\'nski nonlinear electrodynamics with spontaneously Lorentz-breaking magnetic vacua}

\author{C. A. Escobar\thanksref{e1,addr1}
        \and R. Linares\thanksref{e2,addr1}
        \and A. Mart\'in-Ruiz\thanksref{e3,addr2}}

\thankstext{e1}{Corresponding author: carlos.escobar@xanum.uam.mx}
\thankstext{e2}{e-mail: lirr@xanum.uam.mx}
\thankstext{e3}{e-mail: alberto.martin@nucleares.unam.mx}

\institute{Departamento de F\'isica, Universidad Aut\'onoma Metropolitana-Iztapalapa, San Rafael Atlixco 186, 09340 Ciudad de M\'exico, M\'exico\label{addr1}
\and Instituto de Ciencias Nucleares, Universidad Nacional Aut\'onoma de M\'exico, 04510 Ciudad de M\'exico, M\'exico\label{addr2}}

\date{}

\maketitle

\begin{abstract}
We study the Casimir effect in a class of gauge-invariant nonlinear
electrodynamics models designed to admit spontaneously Lorentz-breaking
magnetic vacua. The theory is formulated in the Pleba\'nski
first-order representation, with a single-invariant Hamiltonian
potential \(\widehat V(P)\) as the fundamental nonlinear object. In
this formulation, nontrivial magnetic vacua are stationary points of
the reduced effective Hamiltonian. The symmetry-breaking condition is
controlled by
\(\Sm(P)\equiv\widehat V_P(P)+2P\widehat V_{PP}(P)\), which also
controls the rank of the longitudinal magnetic response and of the
Hamiltonian constraint structure. Taking this Lorentz-breaking
nonlinear electrodynamics as the model under study, we analyze how the
distinction between the regular constant-rank sector and the degenerate
magnetic vacuum enters the parallel-plate Casimir spectrum.
Linearization around a regular magnetic background \(\bar P\), with
\(\Sm(\bar P)\neq0\), yields an ordinary Maxwell-like branch and an
extraordinary anisotropic branch controlled by
\(\alpha(\bar P)=\widehat V_P(\bar P)/\Sm(\bar P)\). We compute the
regularized Casimir energy for magnetic backgrounds perpendicular and
parallel to the plates. In the regular-sector limit
\(\bar P\to P_\star\), with \(\Sm(P_\star)=0\), the extraordinary
branch becomes singular and the parallel-configuration energy diverges.
This divergence is not an infinite physical Casimir force; it signals
that the regular two-branch optical description cannot be continued
uniformly to the rank-changing magnetic vacuum. Direct analysis on the
degenerate surface shows that the extraordinary branch does not survive
as an independent propagating mode for generic momenta. Thus, quantizing
the regular theory and then taking \(\Sm\to0\) is not equivalent to
imposing \(\Sm(P_\star)=0\) before quantization.
\end{abstract}

\section{Introduction}

The Casimir effect is one of the most direct manifestations of the
quantum structure of the electromagnetic vacuum
\cite{Casimir1948,Milton2001}. In its standard form, the confinement of
vacuum fluctuations between conducting boundaries modifies the
zero-point spectrum and produces a measurable force. Because the effect
is determined by the spectrum of admissible field modes, it provides a
sensitive probe of modifications to the underlying electromagnetic
dynamics.

This spectral sensitivity makes the Casimir effect a useful arena for
testing departures from ordinary Lorentz-invariant propagation. If the
field equations contain a preferred direction, either explicitly or
effectively, the allowed mode spectrum can depend not only on the plate
separation, but also on the orientation of the background structure
relative to the boundaries.

Casimir observables have therefore been studied in several
Lorentz-violating field-theory settings. For parallel plates, local and
nonperturbative Green-function approaches show that Lorentz symmetry
breaking can rescale the effective plate separation, modify the vacuum
stress, and introduce orientation-dependent corrections to the Casimir
force
\cite{EscobarMedelMartinRuiz2020,EscobarMartinRuizFrancaGarcia2020,CruzBezerraPetrov2017,CruzBezerraPetrov2018,CruzBezerraMota2020,BezerraCruz2023}.
Related analyses in cylindrical and spherical geometries further show
that the interplay between a preferred direction and the boundary
geometry can become essential
\cite{EscobarRuizCylinder2021,MartinRuizSphere2020}.

Nonlinear electrodynamics (NLED), from the Born--Infeld theory onward,
provides a natural framework in which the electromagnetic vacuum can
behave as an effective medium
\cite{BornInfeld1934,Boillat1970,Novello2000,ObukhovRubilar2002}. In
the presence of background electromagnetic fields, small fluctuations
may propagate anisotropically, as in the classic analysis of photon
propagation in the Euler--Heisenberg vacuum
\cite{BialynickaBirula1970}. This leads to modified dispersion
relations, birefringence, effective optical metrics, and altered mode
densities. Related developments in nonlinear electrodynamics without
birefringence, including Born--Infeld and ModMax-type structures, also
show that nonlinear constitutive properties can strongly constrain the
propagation of electromagnetic fluctuations
\cite{RussoTownsend2023}.

Within this NLED context, two important reference points are the
Euler--Heisenberg effective theory, which gives the standard nonlinear
description of low-energy QED vacuum polarization
\cite{HeisenbergEuler1936,Schwinger1951}, and ModMax electrodynamics,
a conformal and duality-invariant nonlinear extension of Maxwell theory
\cite{Bandos2020}. The Casimir effect in these nonlinear theories was
recently analyzed in Ref.~\cite{Sorge2024}. In that work, the magnetic background is treated as a prescribed
classical configuration around which the theory is linearized. The
resulting Casimir problem is therefore that of fluctuations propagating
in an anisotropic effective medium determined by the chosen background.

The question addressed here is different. We ask what happens to the
Casimir problem when the preferred magnetic direction is not imposed as
an external background, but is tied to a self-sustained magnetic vacuum
of the nonlinear theory itself. This distinction is crucial in
Pleba\'nski nonlinear electrodynamics, because the condition that
selects a nontrivial magnetic vacuum also controls the rank of the
Hamiltonian constraint structure and of the linearized magnetic
constitutive response.

We therefore formulate the theory in the Pleba\'nski first-order
Hamiltonian representation
\cite{Plebanski1970,EscobarPotting2020IJMPA}, using a single-invariant
Hamiltonian potential $\Vhat(P)$ as the fundamental nonlinear object.
In the magnetic branch, the theory admits nontrivial stationary magnetic
vacua that spontaneously break Lorentz symmetry. However, the
fluctuation problem around regular magnetic backgrounds and the
fluctuation problem defined directly at the stationary Lorentz-breaking
magnetic vacuum are not, in general, related by a smooth limiting
procedure.

This point is conceptually related to effective field-theory
descriptions of Lorentz and CPT violation, such as the Standard-Model
Extension
\cite{ColladayKostelecky1997,ColladayKostelecky1998}, where fixed
background tensors encode preferred spacetime directions. In the
present case, however, the anisotropy is not externally prescribed but
emerges from the magnetic branch of the nonlinear Hamiltonian.

Working directly with $\Vhat(P)$, instead of starting from a
conventional Lagrangian $\mathcal L(\mathcal F)$, is not merely a
change of variables. In the Pleba\'nski representation, the reduced
effective Hamiltonian is obtained naturally in terms of the Hamiltonian
variables $\Dvec$ and $\Hvec$, and the conditions for magnetic symmetry
breaking, local stability, and constraint degeneracy can be formulated
within a single Hamiltonian framework. In the magnetic branch, the
relevant combination is \cite{EscobarPotting2020IJMPA}
\begin{equation}
    \Sm(P)
    =
    \Vhat_P(P)+2P\Vhat_{PP}(P).
    \label{eq:intro_Sm}
\end{equation}
Stable Lorentz-breaking magnetic vacua occur at $P=P_\star>0$ with
\begin{equation}
    \Sm(P_\star)=0.
\end{equation}

The same magnetic-branch factor plays several roles at once. It
determines the nontrivial magnetic stationarity condition, governs the
degeneracy of the reduced effective Hamiltonian constraint structure,
and, as we shall show, controls the extraordinary branch of
electromagnetic fluctuations. Its appearance in the strong-field
causality criteria for Pleba\'nski-type nonlinear electrodynamics
further supports the interpretation of $\Sm=0$ as a critical
propagation surface
\cite{Schellstede2016,RussoTownsend2024}. In the present work we use
this observation to identify the boundary of the regular propagation
regime; we do not assume that the single-invariant models considered
below define a causal effective medium across the rank-changing
surface.

The origin of this critical behavior is constitutive. A recent analysis
showed that, at a nontrivial magnetic stationary point, the linearized
magnetic map $\delta\Hvec\mapsto\delta\Bvec$ loses rank in the
longitudinal direction
\cite{EscobarLinares2026Constitutive}. This observation is important
for the Casimir problem. The extraordinary fluctuation branch in the
regular sector probes the inverse of this longitudinal magnetic
response. Therefore, the singular behavior encountered below is not an
infinite direct response of the exact vacuum. It is the singular limit
of a regular inverse-response description continued toward a point
where the underlying constitutive map changes rank.

The central question of this work is therefore how the Casimir spectrum
is affected once this Lorentz-breaking Hamiltonian sector is taken as
the model under study. The key point is that the regular fluctuation
problem is not obtained by evaluating an ordinary anisotropic-medium
Casimir problem at a special value of the magnetic background. It is
defined in a constant-rank sector, whose singular endpoint must be
distinguished from the fluctuation theory formulated directly on the
rank-changing magnetic vacuum.

This requires distinguishing two inequivalent procedures. One may first
quantize the regular two-branch theory and then study the singular
limit $\Sm\to0$. Alternatively, one may impose the exact degeneracy
condition $\Sm(P_\star)=0$ from the outset and quantize the
corresponding critical fluctuation system. These two procedures do not
give the same propagating content, and this noncommutativity is the
main conceptual point of the present analysis.

We proceed in two stages. First, we work in the regular sector,
\begin{equation}
    \Sm(\bar P)\neq0,
\end{equation}
where the fluctuation spectrum contains two propagating branches. We
derive the corresponding dispersion relations and compute the
regularized parallel-plate Casimir energy for magnetic backgrounds
perpendicular and parallel to the plates. Second, we analyze the
singular regular-sector limit associated with the exact
Lorentz-breaking magnetic vacuum, $P_\star$, and show that the
two-branch regular description loses uniform validity because the
inverse longitudinal magnetic response becomes singular. A direct
analysis on the degenerate surface reveals that the extraordinary
branch does not survive as an independent propagating mode for
generic momenta.

The resulting picture is therefore sector dependent. The Casimir
spectrum probes not only the anisotropic dispersion relation of a
regular effective medium, but also the rank of the constitutive and
Hamiltonian structure on which the fluctuation problem is defined.

The paper is organized as follows. Section~\ref{sec:framework_vacua}
introduces the Pleba\'nski Hamiltonian framework and the
spontaneously Lorentz-breaking magnetic vacua. Section~\ref{sec:fluctuations_regular}
derives the electromagnetic fluctuation spectrum in the regular sector.
Section~\ref{sec:casimir_regular} formulates the parallel-plate Casimir
problem and evaluates the regular-sector vacuum energy.
Section~\ref{sec:critical_vacuum} analyzes the singular regular-sector
limit associated with the degenerate Lorentz-breaking magnetic vacuum,
illustrates the mechanism in explicit models, studies the exact
fluctuation spectrum on the degenerate surface, and discusses the range
of validity of the regular optical description.
Section~\ref{sec:conclusions} summarizes the main results and their
conceptual implications. The appendix provides the derivation of the
standing-wave cavity spectrum from the perfect-conductor boundary
conditions.
%%%%%%%%%%%%%%%%%%%%%%%%%%%%%%%%%%%%%%%%%%%%%%%%%%%
\section{Pleba\'nski framework and spontaneously Lorentz-breaking magnetic vacua}
\label{sec:framework_vacua}

\subsection{Pleba\'nski nonlinear electrodynamics}
\label{subsec:plebanski_framework}

We formulate the problem directly within the Pleba\'nski first-order
representation of nonlinear electrodynamics, where the antisymmetric
tensor $P^{\mu\nu}$ and the gauge potential $A_\mu$ are treated as
independent variables. We restrict attention to a single-invariant
sector, in which the nonlinear dynamics is encoded in a structural
Hamiltonian potential $\Vhat(P)$.

The source-free first-order Lagrangian density is
\begin{equation}
\widehat{\mathcal L}(P^{\mu\nu},A_\mu)
=
-P^{\mu\nu}\partial_\mu A_\nu-\Vhat(P),
\label{eq:plebanski_lagrangian}
\end{equation}
with
\begin{equation}
P=\frac14 P_{\mu\nu}P^{\mu\nu}.
\label{eq:P_invariant}
\end{equation}
Introducing the Hamiltonian variables
\begin{equation}
D_i=P^{0i},
\qquad
H_i=-\frac12\epsilon_{ijk}P^{jk},
\label{eq:DH_def}
\end{equation}
one has
\begin{equation}
P=\frac12\left(\mathbf H^2-\mathbf D^2\right).
\label{eq:P_HD}
\end{equation}

Variation with respect to $P^{\mu\nu}$ gives the inverse constitutive
relation
\begin{equation}
F^{\mu\nu}=-\Vhat_P P^{\mu\nu},
\label{eq:inverse_constitutive_tensor}
\end{equation}
with $F^{\mu\nu}=\partial^\mu A^\nu-\partial^\nu A^\mu$. In vector form Eq. (\ref{eq:inverse_constitutive_tensor}) reads
\begin{equation}
\mathbf E=-\Vhat_P\,\mathbf D,
\qquad
\mathbf B=-\Vhat_P\,\mathbf H.
\label{eq:inverse_constitutive_vectors}
\end{equation}
Variation with respect to $A_\mu$, together with the Bianchi identity
for $F_{\mu\nu}$, yields the Maxwell-like equations
\begin{equation}
\nabla\cdot\mathbf D=0,
\qquad
\nabla\times\mathbf H=\partial_t\mathbf D,
\label{eq:maxwell_DH}
\end{equation}
and
\begin{equation}
\nabla\cdot\mathbf B=0,
\qquad
\nabla\times\mathbf E=-\partial_t\mathbf B.
\label{eq:maxwell_EB}
\end{equation}
Thus the theory remains gauge invariant, while the nonlinear
constitutive relations allow for nontrivial electromagnetic vacuum
structure.

After implementation of the second-class constraints, the dynamics is
governed by the reduced effective Hamiltonian density \cite{EscobarPotting2020IJMPA}
\begin{equation}
\mathcal H_{\rm eff}(\mathbf D,\mathbf H)
=
-\mathbf H^2\,\Vhat_P(P)+\Vhat(P),
\label{eq:Heff_DH}
\end{equation}
with $P$ given by Eq.~\eqref{eq:P_HD}. For later convenience we define
\begin{equation}
h=\mathbf H^2,
\qquad
d=\mathbf D^2,
\qquad
P=\frac12(h-d),
\label{eq:hdP_def}
\end{equation}
so that
\begin{equation}
\mathcal H_{\rm eff}(h,d)
=
-h\,\Vhat_P(P)+\Vhat(P).
\label{eq:Heff_hd}
\end{equation}

The relevant point is that the vacuum problem of the model is formulated
at the level of the reduced effective Hamiltonian. The candidate vacuum
configurations are obtained by extremizing $\mathcal H_{\rm eff}$ with
respect to the canonical variables $\mathbf D$ and $\mathbf H$, not by
extremizing the structural potential $\Vhat(P)$ alone. As a result,
magnetic stationarity, Hessian stability, and possible degeneracies of
the constraint structure are all described within a single Hamiltonian
framework. This is the reason why the Pleba\'nski representation is the
appropriate framework for analyzing this particular implementation of
spontaneous Lorentz symmetry breaking.

\subsection{Spontaneously Lorentz-breaking magnetic vacua}
\label{subsec:magnetic_vacua}

We now identify the class of magnetic vacua used in this work to
implement spontaneous Lorentz symmetry breaking. Applying the
stationarity conditions to $\mathcal H_{\rm eff}$, one finds
\begin{equation}
\frac{\partial\mathcal H_{\rm eff}}{\partial D_i}
=
D_i\left(h\Vhat_{PP}-\Vhat_P\right),
\label{eq:dHeff_dD}
\end{equation}
and
\begin{equation}
\frac{\partial\mathcal H_{\rm eff}}{\partial H_i}
=
-H_i\left(\Vhat_P+h\Vhat_{PP}\right).
\label{eq:dHeff_dH}
\end{equation}
Besides the trivial vacuum, the theory may admit a nontrivial purely
magnetic branch characterized by
\begin{equation}
\mathbf D_\star=0,
\qquad
\mathbf H_\star\neq0.
\label{eq:magnetic_branch}
\end{equation}
Defining
\begin{equation}
h_\star=\mathbf H_\star^2,
\qquad
P_\star=\frac{h_\star}{2}>0,
\label{eq:Pstar_def}
\end{equation}
the nontrivial magnetic stationarity condition becomes
\begin{equation}
\Vhat_P(P_\star)+h_\star\Vhat_{PP}(P_\star)
=
\Vhat_P(P_\star)+2P_\star\Vhat_{PP}(P_\star)
=0.
\label{eq:magnetic_stationarity_explicit}
\end{equation}

It is useful to introduce the magnetic-branch function
\begin{equation}
\Sm(P)
\equiv
\Vhat_P(P)+2P\Vhat_{PP}(P).
\label{eq:Sm_def}
\end{equation}
The Lorentz-breaking magnetic vacuum of the model is therefore defined
by
\begin{equation}
\Sm(P_\star)=0,
\qquad
P_\star>0.
\label{eq:magnetic_vacuum_condition}
\end{equation}
Since $\mathbf H_\star$ is a nonzero spatial vector, this vacuum selects
a preferred direction and spontaneously breaks Lorentz symmetry while
preserving gauge invariance.

Thus \(\Sm(P)\) is the control parameter that separates the regular
magnetic sector from the degenerate symmetry-breaking vacuum.

For the symmetry-breaking vacuum to represent a locally stable branch
of the Hamiltonian model, the nonzero Hessian eigenvalues of the reduced
Hamiltonian must be positive. In terms of the variables $(h,d)$, these
conditions are controlled, up to positive numerical factors, by
\cite{EscobarPotting2020IJMPA,PlacidoFlores2026Stable}
\begin{equation}
\mathcal{H}_d>0,
\qquad
\mathcal{H}_{hh}>0.
\label{eq:stability_conditions}
\end{equation}
The first condition is governed by
\begin{equation}
\mathcal{H}_d
=
\frac12\left(h\Vhat_{PP}-\Vhat_P\right).
\label{eq:Hd_def}
\end{equation}
Evaluated at the magnetic vacuum, where
$h_\star\Vhat_{PP}(P_\star)=-\Vhat_P(P_\star)$, this gives
\begin{equation}
\mathcal{H}_d\big|_\star
=
-\Vhat_P(P_\star).
\label{eq:Hd_star}
\end{equation}
Hence local stability implies
\begin{equation}
\Vhat_P(P_\star)<0.
\label{eq:Vp_negative}
\end{equation}
This sign condition will be important for identifying the hyperbolic
real-frequency side of the regular propagation sector adjacent to the
degenerate surface.

The second nontrivial stability condition is
\begin{equation}
\mathcal{H}_{hh}
=
-\frac14
\left(
3\Vhat_{PP}+h\Vhat_{PPP}
\right)>0.
\label{eq:Hhh_def}
\end{equation}
At the magnetic vacuum this becomes
\begin{equation}
\mathcal{H}_{hh}\big|_\star
=
-\frac14
\left[
3\Vhat_{PP}(P_\star)
+
2P_\star\Vhat_{PPP}(P_\star)
\right]>0.
\label{eq:Hhh_star}
\end{equation}
Together with boundedness of the reduced effective Hamiltonian, these
conditions select the locally stable symmetry-breaking branches used in
the explicit examples of Sec.~\ref{sec:critical_vacuum}. They should be
understood as internal consistency conditions of the Hamiltonian models,
not as a separate phenomenological viability claim.

In the following sections we keep this stationary vacuum distinct from
the regular magnetic backgrounds used to define the fluctuation problem.
The latter are constant purely magnetic configurations
\begin{equation}
    \bar{\Dvec}=0,\qquad
    \bar{\Hvec}=\bar H\,\hat{\mathbf n},\qquad
    \bar P=\frac12 \bar H^2 ,
    \label{eq:regular_background}
\end{equation}
with
\begin{equation}
    \Sm(\bar P)\neq0 .
    \label{eq:regular_sector_condition}
\end{equation}
They define the constant-rank sector in which the regular optical modes
and the corresponding Casimir sums are computed. The degenerate vacuum
\(P_\star\), satisfying \(\Sm(P_\star)=0\), is treated separately.

%%%%%%%%%%%%%%%%%%%%%%%%%%%%%%%%%%%%%%%%%%%%%
%%%%%%%%%%%%%%%%%%%%%%%%%%%%%%%%%%%%%%%%%%%%%

\section{Electromagnetic fluctuations in the regular sector}
\label{sec:fluctuations_regular}

\subsection{Linearization around a regular magnetic background}
\label{subsec:linearization_regular}

We now study small electromagnetic fluctuations around a constant
purely magnetic background in the regular sector. This background is not
the degenerate symmetry-breaking vacuum itself; it belongs to the
constant-rank sector in which the fluctuation problem is well defined
with $\Sm(\bar P)\neq0$. We write
\begin{equation}
\bar{\mathbf D}=0,
\qquad
\bar{\mathbf H}=\bar H\,\hat{\mathbf n},
\qquad
\hat{\mathbf n}^{\,2}=1,
\label{eq:regular_background}
\end{equation}
with
\begin{equation}
\bar P=\frac12\bar H^2.
\label{eq:Pbar_def}
\end{equation}

We introduce perturbations according to
\begin{equation}
\mathbf D=\delta\mathbf D,
\qquad
\mathbf H=\bar{\mathbf H}+\delta\mathbf H.
\label{eq:fluctuation_def}
\end{equation}
The background-dependent coefficients that enter the linearized
constitutive relations are
\begin{equation}
p=\Vhat_P(\bar P),
\qquad
u=\Vhat_{PP}(\bar P),
\qquad
s=p+\bar H^2u=\Sm(\bar P).
\label{eq:pus_def}
\end{equation}
Thus the regular sector corresponds to $s\neq0$.

To first order, the variation of the invariant $P$ is
\begin{equation}
\delta P=\bar{\mathbf H}\cdot\delta\mathbf H.
\label{eq:delta_P}
\end{equation}
From the electric constitutive relation
\begin{equation}
\mathbf E=-\Vhat_P\,\mathbf D,
\end{equation}
one obtains
\begin{equation}
\delta\mathbf E=-p\,\delta\mathbf D.
\label{eq:delta_E}
\end{equation}
For the magnetic induction,
\begin{equation}
\mathbf B=-\Vhat_P\,\mathbf H,
\end{equation}
the linear variation gives
\begin{equation}
\delta\mathbf B
=
-p\,\delta\mathbf H
-u(\bar{\mathbf H}\cdot\delta\mathbf H)\bar{\mathbf H}.
\label{eq:delta_B_general}
\end{equation}
Decomposing the magnetic fluctuation into transverse and longitudinal
components with respect to $\hat{\mathbf n}$,
\begin{equation}
\delta\mathbf H
=
\delta\mathbf H_\perp
+
\delta H_\parallel\,\hat{\mathbf n},
\qquad
\delta H_\parallel=\hat{\mathbf n}\cdot\delta\mathbf H,
\label{eq:H_decomposition}
\end{equation}
Eq.~\eqref{eq:delta_B_general} becomes
\begin{equation}
\delta\mathbf B
=
-p\,\delta\mathbf H_\perp
-
s\,\delta H_\parallel\,\hat{\mathbf n}.
\label{eq:delta_B_decomposed}
\end{equation}

Equation~\eqref{eq:delta_B_decomposed} is the key local structure
behind the singular limit. In the regular sector, the magnetic response
map $\delta\mathbf H\mapsto\delta\mathbf B$ is invertible in both the
transverse and longitudinal channels. However, the longitudinal
eigenvalue is precisely $-s=-\Sm(\bar P)$.  Hence, in the limit \(\bar P\to P_\star\) defining the exact magnetic
vacuum,
\begin{equation}
\delta\mathbf B_\parallel
=
-\Sm(\bar P)\,\delta H_\parallel\,\hat{\mathbf n}
\longrightarrow 0.
\label{eq:direct_response_limit}
\end{equation}
The direct longitudinal response therefore does not diverge; it loses
rank. Singular factors arise only when the regular map is inverted
before taking the limit:
\begin{equation}
\delta H_\parallel\,\hat{\mathbf n}
=
-\frac{1}{\Sm(\bar P)}\,\delta\mathbf B_\parallel.
\label{eq:inverse_response_limit}
\end{equation}
Thus the factors of $1/\Sm$ that appear below should be understood as
inverse-response singularities associated with a rank-changing
constitutive map. This is the same mechanism identified in the
constitutive analysis of Hamiltonian degeneracy in
Ref.~\cite{EscobarLinares2026Constitutive}.

The linearized field equations are
\begin{equation}
\nabla\cdot\delta\mathbf D=0,
\qquad
\nabla\times\delta\mathbf H=\partial_t\delta\mathbf D,
\label{eq:lin_maxwell_DH}
\end{equation}
and
\begin{equation}
\nabla\cdot\delta\mathbf B=0,
\qquad
\nabla\times\delta\mathbf E=-\partial_t\delta\mathbf B.
\label{eq:lin_maxwell_EB}
\end{equation}
They define an anisotropic effective propagation problem controlled by
the regular-sector coefficients $p$ and $s$. The extraordinary fluctuation branch becomes singular precisely when
the longitudinal response coefficient \(s\) tends to zero.

\subsection{Plane-wave spectrum}
\label{subsec:plane_wave_spectrum}

We now determine the bulk fluctuation spectrum in the regular sector.
Consider plane-wave perturbations
\begin{equation}
\delta X(\mathbf x,t)
=
\delta X_0\,e^{i(\mathbf k\cdot\mathbf x-\omega t)}.
\label{eq:plane_wave_ansatz}
\end{equation}
The linearized Maxwell equations imply
\begin{equation}
\mathbf k\cdot\delta\mathbf D=0,
\qquad
\mathbf k\times\delta\mathbf H=-\omega\,\delta\mathbf D,
\label{eq:plane_wave_DH}
\end{equation}
and
\begin{equation}
\mathbf k\cdot\delta\mathbf B=0,
\qquad
\mathbf k\times\delta\mathbf E=\omega\,\delta\mathbf B.
\label{eq:plane_wave_EB}
\end{equation}

Because the background preserves axial symmetry around
$\hat{\mathbf n}$, we choose coordinates such that
\begin{equation}
\hat{\mathbf n}=\hat{\mathbf z},
\qquad
\mathbf k=(\kappa,0,q),
\label{eq:k_choice}
\end{equation}
where
\begin{equation}
q\equiv k_\parallel,
\qquad
\kappa\equiv k_\perp .
\end{equation}
Writing
\begin{equation}
\delta\mathbf H=(H_x,H_y,H_z),
\end{equation}
one obtains the algebraic system
\begin{equation}
p(\omega^2-q^2)H_x+pq\kappa H_z=0,
\label{eq:alg_Hx}
\end{equation}
\begin{equation}
p(\omega^2-q^2-\kappa^2)H_y=0,
\label{eq:alg_Hy}
\end{equation}
and
\begin{equation}
p\kappa q H_x+(s\omega^2-p\kappa^2)H_z=0.
\label{eq:alg_Hz}
\end{equation}

The decoupled equation for $H_y$ gives the ordinary branch
\begin{equation}
\omega_1^2=q^2+\kappa^2.
\label{eq:ordinary_branch}
\end{equation}
This mode has the standard Maxwell dispersion relation. Its frequency
is independent of the nonlinear anisotropic coefficient $s$, although
its normalization is still determined by the underlying Hamiltonian
theory.

The coupled $(H_x,H_z)$ sector is described by
\begin{equation}
\begin{pmatrix}
p(\omega^2-q^2) & pq\kappa \\
p\kappa q & s\omega^2-p\kappa^2
\end{pmatrix}
\begin{pmatrix}
H_x\\
H_z
\end{pmatrix}
=0.
\label{eq:coupled_block_regular}
\end{equation}
Its determinant contains a static constrained solution and a nonstatic
propagating branch. The latter is
\begin{equation}
\omega_2^2
=
q^2+\frac{p}{s}\,\kappa^2.
\label{eq:extraordinary_branch_raw}
\end{equation}
Introducing the regular-sector anisotropy parameter
\begin{equation}
\alpha(\bar P)
=
\frac{p}{s}
=
\frac{\Vhat_P(\bar P)}
{\Vhat_P(\bar P)+2\bar P\Vhat_{PP}(\bar P)}
=
\frac{\Vhat_P(\bar P)}{\Sm(\bar P)},
\label{eq:alpha_def}
\end{equation}
the two propagating branches can be written as
\begin{equation}
\omega_1^2=k_\parallel^2+k_\perp^2,
\qquad
\omega_2^2=k_\parallel^2+\alpha\,k_\perp^2.
\label{eq:two_branches_regular}
\end{equation}

The regular optical regime requires real frequencies for arbitrary
transverse momenta, and therefore
\begin{equation}
\alpha>0.
\label{eq:alpha_positive}
\end{equation}
For the locally stable symmetry-breaking branches considered in this
work, $\Vhat_P(P_\star)<0$. Hence, by continuity,
\(p=\Vhat_P(\bar P)\) remains negative in the corresponding regular
sector adjacent to the degenerate surface. The condition
$\alpha=p/s>0$ then selects the hyperbolic real-frequency side of the
regular sector,
\begin{equation}
s=\Sm(\bar P)<0.
\label{eq:physical_side}
\end{equation}
On this real-frequency side,
\begin{equation}
\alpha(\bar P)\to+\infty
\qquad
\text{as}
\qquad
\bar P\to P_\star .
\label{eq:alpha_diverges}
\end{equation}

Thus the ordinary branch remains Maxwell-like, while the extraordinary
branch directly probes the inverse longitudinal magnetic response. Its
singular behavior as $\Sm(\bar P)\to0$ is the first spectral signal
that the regular two-branch description cannot be continued naively to
the exact degenerate vacuum.

\subsection{Polarization structure}
\label{subsec:polarization_structure}

The two branches also have distinct polarization structure. The ordinary
branch is obtained from the decoupled sector,
\begin{equation}
H_x=H_z=0,
\qquad
H_y\neq0.
\label{eq:ordinary_polarization}
\end{equation}
It therefore corresponds to a Maxwell-like polarization transverse to
both the propagation plane and the magnetic background direction.

The extraordinary branch lies in the $(H_x,H_z)$ sector. For
$\kappa\neq0$, Eqs.~\eqref{eq:alg_Hx}--\eqref{eq:alg_Hz} give
\begin{equation}
H_x
=
-\frac{s q}{p\kappa}\,H_z.
\label{eq:extraordinary_polarization}
\end{equation}
This is the second propagating polarization of the regular theory. Its
existence relies on the invertibility of the longitudinal magnetic
response, and its dispersion depends explicitly on the ratio
$p/s=\alpha$.

The collinear sector $\kappa=0$ is exceptional because the distinction
between transverse and longitudinal propagation relative to
$\hat{\mathbf n}$ degenerates. In the regular bulk theory this
lower-dimensional sector can be treated separately, but it does not
affect the Casimir mode density considered below. The same exceptional
sector will reappear when the exact degenerate surface is analyzed in
Sec.~\ref{sec:critical_vacuum}.

%%%%%%%%%%%%
%%%%%%%%%%%%
%%%%%%%%%%%%

\section{Casimir effect in the regular propagation regime}
\label{sec:casimir_regular}

\subsection{Parallel-plate geometry}
\label{subsec:parallel_plate_geometry}

We now formulate the parallel-plate Casimir problem for the
regular-sector fluctuation spectrum derived in
Sec.~\ref{sec:fluctuations_regular}. The conducting plates are located
at
\begin{equation}
    z=0,
    \qquad
    z=a.
    \label{eq:plates_location}
\end{equation}
Throughout this section, the calculation is performed around a regular
purely magnetic background $\bar P$ satisfying
\begin{equation}
    \Sm(\bar P)\neq0.
    \label{eq:regular_sector_casimir}
\end{equation}
Thus, the mode sums below define the Casimir problem in the regular
constant-rank sector of the model. They should not be identified with
the fluctuation problem obtained after imposing the degenerate vacuum
condition $\Sm(P_\star)=0$. That problem is addressed separately in
Sec.~\ref{sec:critical_vacuum}.

For a perfect conductor, the boundary conditions are imposed on the
physical electromagnetic fields,
\begin{equation}
\hat{\mathbf z}\times \delta\mathbf E=0,
\qquad
\hat{\mathbf z}\cdot \delta\mathbf B=0,
\qquad
z=0,a .
\label{eq:perfect_conductor_bc}
\end{equation}
Since the fundamental variables in the Pleba\'nski formulation are
$\delta\mathbf D$ and $\delta\mathbf H$, the translation of
Eq.~\eqref{eq:perfect_conductor_bc} into cavity mode conditions is not
automatic. In~\ref{app:cavity_modes} we verify that, for the
two orientations considered below, the normal momentum is quantized as
\begin{equation}
k_z=\frac{n\pi}{a},
\qquad
n=1,2,\ldots,
\label{eq:kz_quantization}
\end{equation}
while the momentum components parallel to the plates remain continuous.
The $n=0$ sector does not contribute to the separation-dependent part
of the regularized vacuum energy relevant for the force and is omitted
from the finite Casimir contribution.

The extraordinary dispersion relation,
\begin{equation}
    \omega_2^2
    =
    k_\parallel^2+\alpha k_\perp^2,
    \label{eq:omega2_casimir_general}
\end{equation}
is defined relative to the preferred direction selected by the regular
magnetic background, not relative to the cavity normal. Therefore the
Casimir spectrum depends on the orientation of $\bar{\mathbf H}$ with
respect to the conducting plates.

In this section we write
\begin{equation}
    \alpha\equiv\alpha(\bar P)
    =
    \frac{\Vhat_P(\bar P)}{\Sm(\bar P)},
    \qquad
    \alpha>0,
    \label{eq:alpha_casimir}
\end{equation}
where the positivity condition specifies the real-frequency regular
sector in which the Casimir mode sum is defined.
The background is kept fixed when the plate separation is varied, so
$\alpha$ is treated as independent of $a$ in the computation of the
force.

The formal zero-point energy per unit area is
\begin{equation}
    \frac{E_0}{A}
    =
    \frac12
    \sum_{r=1}^{2}
    \sum_{n=1}^{\infty}
    \int
    \frac{\dd^2k}{(2\pi)^2}\,
    \omega_{r,n},
    \label{eq:zero_point}
\end{equation}
where $r=1,2$ labels the ordinary and extraordinary propagating
branches in the regular sector. This expression is ultraviolet
divergent; therefore, we use zeta-function regularization and retain only the
finite geometry-dependent part after subtracting contributions
independent of the plate separation.

For reference, a single Maxwell-like polarization contributes
\begin{equation}
    \frac{E_{\rm one}}{A}
    =
    -\frac{\pi^2}{1440a^3},
    \label{eq:E_one_result}
\end{equation}
so that the standard Maxwell result is
\begin{equation}
    \frac{E_{\rm Maxwell}}{A}
    =
    -\frac{\pi^2}{720a^3}.
    \label{eq:E_Maxwell_result}
\end{equation}

\subsection{Regular magnetic background perpendicular to the plates}
\label{subsec:perpendicular_background}

We first consider a regular magnetic background oriented normal to the
cavity,
\begin{equation}
    \bar{\mathbf H}=\bar H\,\hat{\mathbf z}.
    \label{eq:Hbar_perp}
\end{equation}
In this configuration, the preferred direction coincides with the
normal to the plates, and therefore
\begin{equation}
    k_\parallel=k_z=\frac{n\pi}{a},
    \qquad
    k_\perp^2=k_x^2+k_y^2.
    \label{eq:k_perp_geometry}
\end{equation}
The ordinary and extraordinary frequencies are
\begin{equation}
    \omega_{1,n}^{\perp}
    =
    \sqrt{
    k_x^2+k_y^2+\left(\frac{n\pi}{a}\right)^2
    },
    \label{eq:omega1_perp}
\end{equation}
and
\begin{equation}
    \omega_{2,n}^{\perp}
    =
    \sqrt{
    \left(\frac{n\pi}{a}\right)^2
    +
    \alpha(k_x^2+k_y^2)
    }.
    \label{eq:omega2_perp}
\end{equation}
Thus, in the perpendicular geometry, the anisotropy rescales only the
continuous momentum components parallel to the plates.

The ordinary branch contributes the one-polarization Maxwell result,
\begin{equation}
    \frac{E_1^\perp}{A}
    =
    -\frac{\pi^2}{1440a^3}.
    \label{eq:E1_perp}
\end{equation}
For the extraordinary branch, define
\begin{equation}
    q_x=\sqrt{\alpha}\,k_x,
    \qquad
    q_y=\sqrt{\alpha}\,k_y.
    \label{eq:rescale_perp}
\end{equation}
The frequency becomes isotropic in the variables $(q_x,q_y)$, while the
measure transforms as
\begin{equation}
    \dd k_x\,\dd k_y
    =
    \frac{1}{\alpha}\,
    \dd q_x\,\dd q_y.
    \label{eq:jacobian_perp}
\end{equation}
Hence the finite contribution of the extraordinary branch is
\begin{equation}
    \frac{E_2^\perp}{A}
    =
    -\frac{\pi^2}{1440a^3}\frac{1}{\alpha}.
    \label{eq:E2_perp}
\end{equation}
The total regular-sector Casimir energy is therefore
\begin{equation}
    \frac{E_{\rm Cas}^{\perp}}{A}
    =
    -\frac{\pi^2}{1440a^3}
    \left(
    1+\frac{1}{\alpha}
    \right).
    \label{eq:E_perp_total}
\end{equation}
The corresponding force per unit area is
\begin{equation}
    \frac{F_{\rm Cas}^{\perp}}{A}
    =
    -\frac{\partial}{\partial a}
    \left(
    \frac{E_{\rm Cas}^{\perp}}{A}
    \right)
    =
    -\frac{\pi^2}{480a^4}
    \left(
    1+\frac{1}{\alpha}
    \right).
    \label{eq:F_perp}
\end{equation}
In this orientation, the anisotropy modifies only the continuum
momentum measure. Consequently, the perpendicular regular-sector energy
remains finite in the critical limit $\alpha\to+\infty$.

%%%%%%%%%%%%%%%%%%%%%%%%%%%%%%%%
%%%

\subsection{Regular magnetic background parallel to the plates}
\label{subsec:parallel_background}

We now consider the complementary configuration in which the regular
magnetic background lies parallel to the conducting plates. Without loss
of generality, we choose
\begin{equation}
    \bar{\mathbf H}=\bar H\,\hat{\mathbf x}.
    \label{eq:Hbar_parallel}
\end{equation}
Then
\begin{equation}
    k_\parallel=k_x,
    \qquad
    k_\perp^2
    =
    k_y^2
    +
    \left(\frac{n\pi}{a}\right)^2.
    \label{eq:k_parallel_geometry}
\end{equation}
The corresponding frequencies are
\begin{equation}
    \omega_{1,n}^{\parallel}
    =
    \sqrt{
    k_x^2+k_y^2+\left(\frac{n\pi}{a}\right)^2
    },
    \label{eq:omega1_parallel}
\end{equation}
and
\begin{equation}
    \omega_{2,n}^{\parallel}
    =
    \sqrt{
    k_x^2
    +
    \alpha
    \left[
    k_y^2+\left(\frac{n\pi}{a}\right)^2
    \right]
    }.
    \label{eq:omega2_parallel}
\end{equation}
Unlike the perpendicular case, the anisotropy now acts directly on the
discretized cavity momentum.

The ordinary branch again gives
\begin{equation}
    \frac{E_1^\parallel}{A}
    =
    -\frac{\pi^2}{1440a^3}.
    \label{eq:E1_parallel}
\end{equation}
For the extraordinary branch, introduce
\begin{equation}
    q_y=\sqrt{\alpha}\,k_y,
    \qquad
    a_{\rm eff}=\frac{a}{\sqrt{\alpha}}.
    \label{eq:rescale_parallel}
\end{equation}
The frequency then takes the standard isotropic form with effective
plate separation $a_{\rm eff}$, while the momentum measure contributes
a Jacobian factor $1/\sqrt{\alpha}$. Therefore
\begin{equation}
    \frac{E_2^\parallel}{A}
    =
    \frac{1}{\sqrt{\alpha}}
    \left[
    -\frac{\pi^2}{1440a_{\rm eff}^3}
    \right]
    =
    -\frac{\pi^2}{1440a^3}\,\alpha.
    \label{eq:E2_parallel}
\end{equation}
The full regular-sector Casimir energy is
\begin{equation}
    \frac{E_{\rm Cas}^{\parallel}}{A}
    =
    -\frac{\pi^2}{1440a^3}
    \left(1+\alpha\right),
    \label{eq:E_parallel_total}
\end{equation}
with corresponding force
\begin{equation}
    \frac{F_{\rm Cas}^{\parallel}}{A}
    =
    -\frac{\partial}{\partial a}
    \left(
    \frac{E_{\rm Cas}^{\parallel}}{A}
    \right)
    =
    -\frac{\pi^2}{480a^4}
    \left(1+\alpha\right).
    \label{eq:F_parallel}
\end{equation}

The difference between Eqs.~\eqref{eq:E_perp_total} and
\eqref{eq:E_parallel_total} is geometric. When the magnetic background
is perpendicular to the plates, the anisotropy rescales only the
continuous momentum measure. When it is parallel to the plates, it also
rescales the discrete cavity spacing. This is why the parallel
configuration develops the divergent regular-sector contribution in the
critical limit $\alpha(\bar P)\to+\infty$, analyzed in the next
section.

%%%%%%%%%%%%%%%%%%%%%%%%%%%%
%%%%%%%%%%%%%%%%%%%%%%%%%%%%

\section{Singular regular-sector limit and degenerate magnetic vacuum}
\label{sec:critical_vacuum}

\subsection{Breakdown of the regular-sector limit}
\label{subsec:breakdown_regular}

The Casimir energies obtained in Sec.~\ref{sec:casimir_regular} were
derived entirely within the regular propagation sector, where
\begin{equation}
\Sm(\bar P)\neq0,
\end{equation}
and the fluctuation spectrum contains two propagating branches. We now
examine what this regular-sector result does when the background is
formally taken toward the Lorentz-breaking magnetic vacuum,
\begin{equation}
\bar P\to P_\star,
\qquad
\Sm(P_\star)=0.
\end{equation}

Using the regular-sector Casimir energies obtained in
Eqs.~\eqref{eq:E_perp_total} and \eqref{eq:E_parallel_total}, this
limit gives
\begin{equation}
\frac{E_{\rm Cas}^{\perp}}{A}
=
-\frac{\pi^2}{1440a^3}
\left(
1+\frac{1}{\alpha}
\right)
\longrightarrow
-\frac{\pi^2}{1440a^3},
\label{eq:perp_critical_limit}
\end{equation}
whereas
\begin{equation}
\frac{E_{\rm Cas}^{\parallel}}{A}
=
-\frac{\pi^2}{1440a^3}
(1+\alpha)
\longrightarrow
-\infty.
\label{eq:parallel_critical_divergence}
\end{equation}

\bigskip

The divergence in the parallel configuration identifies the
regular-sector limit as singular. This is also evident from the
extraordinary dispersion relation in Eq.~\eqref{eq:omega2_casimir_general}:
as \(\alpha\to+\infty\), regular-sector extraordinary fluctuations with
nonzero transverse momentum are pushed to parametrically large
frequency. In the present interpretation, this is what follows when the NLED model
is adopted as the system under study and its regular Casimir sector is
driven toward the corresponding rank-changing surface. The answer is that the regular two-branch
optical description cannot be evaluated directly at the degenerate
vacuum. The corresponding sector distinction is discussed next.

\subsection{Rank-changing constrained dynamics and sector interpretation}
\label{subsec:rank_changing_sectors}

The apparent behavior in the parallel configuration should not be read
as a transition of one and the same Casimir energy from infinity to a
finite value. Rather, two different sector-dependent quantities are being
compared: the singular endpoint of the regular two-branch expression
and the zero-point contribution obtained from the fluctuation spectrum
formulated directly on the degenerate surface. Thus,
\begin{equation}
\lim_{\bar P\to P_\star} E_{\rm Cas}^{\rm reg}(\bar P)
\neq
E_{\rm Cas}^{\rm deg}(P_\star) .
\label{eq:casimir_sector_noncommutativity}
\end{equation}
The limit \(\bar P\to P_\star\) can be studied as a singular
diagnostic of the regular sector, but it should not be identified with
the fluctuation theory obtained after imposing
\(S_{\rm m}(P_\star)=0\) from the outset. The point
\(S_{\rm m}(P_\star)=0\) is a Hamiltonian-degenerate surface where the
longitudinal magnetic constitutive eigenvalue vanishes and the
corresponding constrained structure changes rank.

The origin of the singular regular limit can be seen from the
linearized magnetic response. As displayed in
Sec.~\ref{subsec:linearization_regular},
\begin{equation}
\delta\mathbf B
=
-p\,\delta\mathbf H_\perp
-
S_{\rm m}(\bar P)\,\delta H_\parallel\,\hat{\mathbf n}.
\label{eq:rank_response_sector}
\end{equation}
For \(S_{\rm m}(\bar P)\neq0\), the longitudinal channel is invertible,
and its inverse enters the extraordinary branch,
\begin{equation}
\delta H_\parallel\,\hat{\mathbf n}
=
-\frac{1}{S_{\rm m}(\bar P)}\,\delta\mathbf B_\parallel .
\label{eq:rank_inverse_sector}
\end{equation}
Accordingly,
\begin{equation}
\alpha(\bar P)
=
\frac{\widehat V_P(\bar P)}{S_{\rm m}(\bar P)}
\label{eq:alpha_rank_sector}
\end{equation}
is a regular-sector inverse-response parameter. For every fixed
background with \(S_{\rm m}(\bar P)\neq0\), the quadratic fluctuation
problem is well defined. What fails as \(\bar P\to P_\star\) is the
uniformity of this regular expansion: the inverse longitudinal
response, and hence the extraordinary branch constructed from it,
becomes singular. The direct constitutive response does not become
infinite; it loses rank at \(S_{\rm m}(P_\star)=0\), where the
longitudinal channel satisfies
\(\delta\mathbf B_\parallel=0\) for finite \(\delta H_\parallel\).

This interpretation is supported by the Hamiltonian structure of
Pleba\'nski nonlinear electrodynamics. The Dirac analysis of magnetic
Lorentz-breaking vacua shows that the same stationarity condition that
selects the nontrivial magnetic vacuum also makes the relevant
second-class-constraint matrix degenerate
\cite{EscobarPotting2020IJMPA}. The constitutive origin of this result
was clarified in Ref.~\cite{EscobarLinares2026Constitutive}: the
magnetic constitutive Jacobian appears as a local block of the
Poisson-bracket matrix of second-class constraints, so that the loss of
rank of the map \(\delta\mathbf H\mapsto\delta\mathbf B\) is inherited
by the Hamiltonian constraint matrix. Thus the limit
\(\alpha\to+\infty\) reflects a rank-changing constrained system,
rather than an ordinary smooth limit of an anisotropic optical medium.

This type of nonuniform limit is familiar in field-theoretic systems in
which a coefficient controlling the quadratic dynamics or the constraint
classification vanishes on special backgrounds. Examples include
ghost-condensate effective theories, Lifshitz points, partially
massless higher-spin fields, canonical bifurcation in higher-derivative
constrained systems, and several implementations of spontaneous
Lorentz breaking through nonlinear or constrained structures
\cite{ArkaniHamed2004GhostCondensate,Hornreich1975Lifshitz,DeserWaldron2001PartialMasslessness,DeserErtlGrumiller2013,BluhmGagnePottingVrublevskis2008,BonderEscobar2016,Seifert2019SingularHamiltonians,EscobarLinares2022}. The common
structural lesson is that the limiting regular theory and the theory
defined exactly on the critical surface need not have the same
propagating content.

This sector distinction is also consistent with the standard
assumptions of the Hamiltonian theory of constrained systems. The
regular Dirac construction assumes regular constraints and a constant
rank of the relevant constraint matrix on the sector under
consideration. In particular, the construction of the Dirac bracket
requires the inverse of the matrix formed by the Poisson brackets among
the second-class constraints. Hence, when this matrix changes rank, the
degenerate surface should not be regarded as an ordinary point of the
same regular canonical description \cite{HenneauxTeitelboim1992}. The
relevant distinction in the present problem is therefore
\begin{equation}
S_{\rm m}(\bar P)\neq0
\qquad\text{versus}\qquad
S_{\rm m}(P_\star)=0 .
\label{eq:sector_distinction}
\end{equation}
The first case is the regular constant-rank sector, where the
two-branch fluctuation spectrum is well defined. The second is the
rank-changing degenerate surface, where the fluctuation problem must be
formulated after imposing \(S_{\rm m}(P_\star)=0\). This is also the
lesson of irregular and degenerate Hamiltonian systems: when the rank
of the symplectic or constraint matrix changes, one either regularizes
the constraints when possible or restricts the canonical analysis to
sectors of constant rank; in irreducible degenerate systems, the
degeneracy surface may separate dynamically distinct sectors
\cite{MiskovicZanelli2003,SaavedraTroncosoZanelli2001,MiskovicTroncosoZanelli2005,DeMicheliZanelli2012}.

Consequently, the present calculation does not model a dynamical
crossing through \(S_{\rm m}=0\). It compares the regular
constant-rank sector with the fluctuation problem formulated directly
on the rank-changing surface. The latter case is analyzed explicitly
in Sec.~\ref{subsec:exact_degenerate_spectrum}.

Finally, the other side of the degenerate surface does not define the
same regular Casimir problem. For the locally stable symmetry-breaking
branches considered here, continuity of \(\widehat V_P\) in the regular
sector gives \(p=\widehat V_P(\bar P)<0\). The real-frequency regular
side is therefore selected by \(S_{\rm m}(\bar P)<0\), so that
\(\alpha=p/S_{\rm m}>0\). If one formally considers backgrounds on the
side \(S_{\rm m}(\bar P)>0\), then \(\alpha<0\) and the extraordinary
dispersion relation
\begin{equation}
\omega_2^2=k_\parallel^2+\alpha k_\perp^2
\label{eq:negative_alpha_dispersion}
\end{equation}
becomes negative for sufficiently transverse momenta. This is a loss of
hyperbolicity, or a gradient instability, not a well-defined regular
Casimir regime. Consequently, the expression obtained for
\(\alpha>0\) cannot be analytically continued through \(S_{\rm m}=0\)
to infer a repulsive force. This restriction is consistent with the
general role of hyperbolicity in nonlinear electrodynamics
\cite{Abalos2015NLEDHyperbolic}.

%%%%%%%%%%%%%%%%%%%%%%%%%%%%%%%%%
\subsection{Rational asymmetric model}
\label{subsec:rational_model}

We now illustrate the mechanism in explicit nonlinear Hamiltonian
potentials constructed to admit locally stable Lorentz-breaking
magnetic branches. The following examples were shown in
Ref.~\cite{PlacidoFlores2026Stable} to possess parameter regions in
which boundedness of the reduced Hamiltonian, magnetic stationarity,
and Hessian stability coexist. We use them here as concrete theoretical
realizations of spontaneous Lorentz symmetry breaking in NLED, rather
than as claims of complete phenomenological viability.

The first example is the rational asymmetric model
\begin{equation}
\Vhat(P)
=
-P+\frac{\lambda P^3}{1+\eta P^2},
\label{eq:rat_potential}
\end{equation}
where $\lambda$ and $\eta$ are real parameters. Its first two
derivatives are
\begin{equation}
\Vhat_P
=
-1+
\lambda P^2
\frac{3+\eta P^2}{(1+\eta P^2)^2},
\label{eq:rat_vp}
\end{equation}
and
\begin{equation}
\Vhat_{PP}
=
-\frac{2\lambda P(\eta P^2-3)}
{(1+\eta P^2)^3}.
\label{eq:rat_vpp}
\end{equation}
Therefore, the magnetic-branch function
\begin{equation}
\Sm(P)=\Vhat_P(P)+2P\Vhat_{PP}(P)
\end{equation}
is
\begin{equation}
\Sm(P)
=
-\frac{
1+(3\eta-15\lambda)P^2
+3\eta^2P^4
+\eta^2(\eta-\lambda)P^6
}
{(1+\eta P^2)^3}.
\label{eq:rat_sm}
\end{equation}

The magnetic Lorentz-breaking vacuum is determined by
\begin{equation}
\Sm(P_\star^{\rm rat})=0,
\qquad
P_\star^{\rm rat}=\frac{h_\star^{\rm rat}}{2}>0.
\label{eq:rat_vacuum_condition}
\end{equation}
Equivalently, in terms of $h_\star^{\rm rat}$, the stationarity
condition becomes
\begin{equation}
\begin{aligned}
0={}&
(\eta^3-\eta^2\lambda)(h_\star^{\rm rat})^6
+12\eta^2(h_\star^{\rm rat})^4
\\
&+(48\eta-240\lambda)(h_\star^{\rm rat})^2
+64 .
\end{aligned}
\label{eq:rat_hstar_equation}
\end{equation}
Thus, the Lorentz-breaking magnetic branch corresponds to the positive
real root of Eq.~\eqref{eq:rat_hstar_equation} lying in the locally
stable parameter domain.

The reduced Hamiltonian is bounded from below for
\begin{equation}
\eta>0,
\qquad
\eta\geq \frac{9}{8}\lambda,
\label{eq:rat_bounded}
\end{equation}
while coexistence of boundedness, magnetic spontaneous Lorentz
breaking, and Hessian stability requires
\begin{equation}
\frac{9}{8}\lambda
\leq
\eta
<
\left(
\frac{25}{16}
+
\frac{5\sqrt5}{16}
\right)\lambda .
\label{eq:rat_stable_region}
\end{equation}
In this region the stationary point is a locally stable magnetic vacuum
of the reduced Hamiltonian system.

The regular-sector anisotropy parameter is
\begin{equation}
\alpha_{\rm rat}(P)
=
\frac{\Vhat_P(P)}{\Sm(P)}.
\end{equation}
Using Eqs.~\eqref{eq:rat_vp} and \eqref{eq:rat_sm}, one obtains
\begin{equation}
\alpha_{\rm rat}(P)
=
\frac{
\left[
1+(2\eta-3\lambda)P^2+\eta(\eta-\lambda)P^4
\right]
(1+\eta P^2)
}
{
1+(3\eta-15\lambda)P^2
+3\eta^2P^4
+\eta^2(\eta-\lambda)P^6
}.
\label{eq:alpha_rat}
\end{equation}
The denominator is precisely the numerator of the degeneracy function
in Eq.~\eqref{eq:rat_sm}. Hence, in the hyperbolic real-frequency limit
\(P\to P_\star^{\rm rat}\),
\begin{equation}
P\to P_\star^{\rm rat},
\qquad
\Sm(P)\to0,
\qquad
\alpha_{\rm rat}(P)\to+\infty.
\label{eq:alpha_rat_diverges}
\end{equation}
The extraordinary branch therefore becomes singular at the same surface
where the Hamiltonian constraint structure degenerates. This provides a
concrete stable realization of the critical Casimir behavior described
above.

\subsection{Logarithmic model}
\label{subsec:logarithmic_model}

As a second example, consider the logarithmic Hamiltonian potential
\begin{equation}
\Vhat(P)
=
-P-\frac{\lambda}{\eta}\ln(1+\eta P),
\label{eq:log_potential}
\end{equation}
with model domain
\begin{equation}
1+\eta P>0.
\label{eq:log_domain}
\end{equation}
The derivatives are
\begin{equation}
\Vhat_P
=
-1-\frac{\lambda}{1+\eta P}
=
-\frac{1+\eta P+\lambda}{1+\eta P},
\label{eq:log_vp}
\end{equation}
and
\begin{equation}
\Vhat_{PP}
=
\frac{\lambda\eta}{(1+\eta P)^2}.
\label{eq:log_vpp}
\end{equation}
Therefore,
\begin{equation}
\Sm(P)
=
\Vhat_P+2P\Vhat_{PP}
=
-\frac{
(\eta P)^2-(\lambda-2)\eta P+(\lambda+1)
}
{(1+\eta P)^2}.
\label{eq:log_sm}
\end{equation}

The magnetic vacuum is determined by
\begin{equation}
\Sm(P_\star^{\log})=0,
\end{equation}
or
\begin{equation}
(\eta P_\star^{\log})^2
-(\lambda-2)\eta P_\star^{\log}
+(\lambda+1)=0.
\label{eq:log_quadratic}
\end{equation}
The two formal roots are
\begin{equation}
\eta P_\pm
=
\frac{\lambda-2\pm\sqrt{\lambda(\lambda-8)}}{2}.
\label{eq:log_roots}
\end{equation}
The branch compatible with local stability is the larger positive root,
\begin{equation}
\begin{aligned}
P_\star^{\log}
&=
\frac{\lambda-2+\sqrt{\lambda(\lambda-8)}}{2\eta},
\\
h_\star^{\log}
&=
2P_\star^{\log}
=
\frac{\lambda-2+\sqrt{\lambda(\lambda-8)}}{\eta}.
\end{aligned}
\label{eq:log_pstar}
\end{equation}

The reduced Hamiltonian is bounded from below for
\begin{equation}
\lambda\geq0,
\qquad
\eta>0.
\label{eq:log_bounded}
\end{equation}
The stronger conditions for boundedness, nontrivial magnetic
stationarity, and Hessian stability select the region
\begin{equation}
\lambda>8,
\qquad
\eta>0.
\label{eq:log_stable_region}
\end{equation}
In this domain, Eq.~\eqref{eq:log_pstar} gives a real positive magnetic
vacuum and the logarithmic argument remains in the model domain.

The corresponding anisotropy parameter is
\begin{equation}
\alpha_{\log}(P)
=
\frac{\Vhat_P(P)}{\Sm(P)}
=
\frac{
(1+\eta P+\lambda)(1+\eta P)
}
{
(\eta P)^2-(\lambda-2)\eta P+(\lambda+1)
}.
\label{eq:alpha_log}
\end{equation}
The denominator vanishes at the magnetic degeneracy surface. Therefore,
on the hyperbolic real-frequency side of the regular sector,
\begin{equation}
P\to P_\star^{\log},
\qquad
\Sm(P)\to0,
\qquad
\alpha_{\log}(P)\to+\infty.
\label{eq:alpha_log_diverges}
\end{equation}
The logarithmic model thus gives a second locally stable realization of
the same critical mechanism: the extraordinary regular-sector mode
becomes singular precisely at the magnetic Lorentz-breaking vacuum.

The rational and logarithmic examples show that the singular behavior
of the regular-sector Casimir spectrum is not tied to a special choice
of \(\widehat V(P)\). It occurs in NLED models whose
Hamiltonian structure is explicitly designed to contain locally stable
magnetic branches that spontaneously break Lorentz symmetry. The
divergence of \(\alpha\) is therefore a structural consequence of the
Hamiltonian-degenerate condition \(\Sm=0\), rather than a pathology of a
particular choice of \(\widehat V(P)\).

\subsection{Exact fluctuation spectrum on the degenerate surface}
\label{subsec:exact_degenerate_spectrum}

We now impose the magnetic degeneracy condition from the outset and
analyze the fluctuation spectrum directly at the exact
Lorentz-breaking vacuum. At the critical surface,
\begin{equation}
\Sm(P_\star)
=
\Vhat_P(P_\star)+2P_\star\Vhat_{PP}(P_\star)
=
0.
\label{eq:Sm_exact_surface}
\end{equation}
The electric response remains
\begin{equation}
\delta\mathbf E
=
-p_\star\,\delta\mathbf D,
\qquad
p_\star=\Vhat_P(P_\star)<0,
\label{eq:delta_E_exact}
\end{equation}
while the magnetic response reduces to
\begin{equation}
\delta\mathbf B
=
-p_\star\,\delta\mathbf H_\perp.
\label{eq:delta_B_exact}
\end{equation}
The longitudinal magnetic response is absent. This is the crucial
difference between the regular and exact degenerate theories: the
longitudinal channel that is invertible for $\Sm(\bar P)\neq0$ becomes
a zero-response direction at $\Sm(P_\star)=0$.

Choose the magnetic vacuum direction as
\begin{equation}
\hat{\mathbf n}=\hat{\mathbf z},
\qquad
\mathbf k=(\kappa,0,q),
\label{eq:exact_k_choice}
\end{equation}
where $q=k_\parallel$ and $\kappa=k_\perp$. Writing
\begin{equation}
\delta\mathbf H=(H_x,H_y,H_z),
\label{eq:exact_H_components}
\end{equation}
the ordinary sector remains
\begin{equation}
p_\star(\omega^2-q^2-\kappa^2)H_y=0.
\label{eq:exact_ordinary_block}
\end{equation}
Since $p_\star\neq0$, this gives the Maxwell-like dispersion relation
\begin{equation}
\omega^2=q^2+\kappa^2.
\label{eq:ordinary_exact}
\end{equation}

The important point is that the degenerate theory does not contain the
same two propagating oscillators as the regular theory. In the regular
sector, the decoupled field $H_y$ gives the ordinary branch, while the
coupled $(H_x,H_z)$ sector gives the extraordinary branch. On the
degenerate surface, however, the longitudinal magnetic response has
lost rank. Therefore the coupled sector must be reanalyzed after
imposing $\Sm(P_\star)=0$, rather than obtained by assigning a limiting
frequency to the regular extraordinary mode.

The would-be extraordinary sector is described by the degenerate
coupled block
\begin{equation}
\begin{pmatrix}
p_\star(\omega^2-q^2) & p_\star q\kappa \\
p_\star\kappa q & -p_\star\kappa^2
\end{pmatrix}
\begin{pmatrix}
H_x\\
H_z
\end{pmatrix}
=0.
\label{eq:exact_coupled_block}
\end{equation}
Its determinant is
\begin{equation}
\det
=
-p_\star^{\,2}\kappa^2\omega^2.
\label{eq:exact_determinant}
\end{equation}
Thus, for generic momenta with $\kappa\neq0$, a nontrivial solution of
this block requires $\omega=0$. The coupled $(H_x,H_z)$ sector therefore
does not supply a nonstatic propagating branch on the degenerate
surface. It is either static or constrained, and hence it does not
define an independent harmonic oscillator contributing a zero-point
energy $\frac12\omega$ to the Casimir sum.

By contrast, the $H_y$ sector remains nondegenerate and gives the
ordinary Maxwell-like dispersion relation in
Eq.~\eqref{eq:ordinary_exact}. Consequently, after imposing
$\Sm(P_\star)=0$ from the outset, the generic propagating spectrum
contains only the ordinary branch. The extraordinary branch that exists
for $\Sm(\bar P)\neq0$ is removed by the rank change of the magnetic
constitutive map.

There is a special collinear sector with $\kappa=0$, where the above
rank argument becomes degenerate. Such modes form a lower-dimensional
subset of momentum space and do not reconstruct the full extraordinary
branch. In particular, they do not affect the Casimir mode density in
the parallel-plate configurations considered here.

The comparison with the regular spectrum therefore gives the explicit
realization of the sector interpretation discussed in
Sec.~\ref{subsec:rank_changing_sectors}. In the regular sector
\(\Sm(\bar P)\neq0\), the ordinary and extraordinary branches are both
present. After imposing \(\Sm(P_\star)=0\) from the outset, the generic
nonstatic spectrum contains only the ordinary branch. Thus the limiting
regular spectrum and the spectrum of the exactly degenerate theory are
not the same.

Equivalently,
\begin{equation}
\lim_{\bar P\to P_\star}
\mathcal Q_{\rm reg}\!\left[\Sm(\bar P)\neq0\right]
\neq
\mathcal Q_{\rm deg}\!\left[\Sm(P_\star)=0\right] .
\label{eq:noncommutativity}
\end{equation}
The divergent regular-sector Casimir energy is therefore a diagnostic
of this noncommutativity, not the zero-point energy associated with the generic propagating spectrum
of the exact Lorentz-breaking state.

\subsection{Physical interpretation and scope of the analysis}
\label{subsec:range_validity}

The purpose of the calculation is not to establish the full
phenomenological viability of the particular single-invariant
Hamiltonian models considered here. Rather, these models are used as
gauge-invariant NLED realizations of spontaneous Lorentz symmetry
breaking. Once such a model is adopted as the system under study, the
Casimir calculation probes how its fluctuation sectors contribute to
the zero-point spectrum.

The singular behavior found above is therefore a sector-dependent property of
the model. For every fixed regular background with
\(\Sm(\bar P)\neq0\) and \(\alpha>0\), the quadratic fluctuation
problem and the corresponding regular-sector Casimir sums are well
defined. What fails in the limit \(\bar P\to P_\star\) is the uniform
continuation of this two-branch description to the rank-changing
surface. Imposing \(\Sm(P_\star)=0\) from the outset gives a different
constrained fluctuation problem, in which the generic nonstatic
extraordinary oscillator is absent.

The side with \(\alpha<0\) does not define an alternative regular
Casimir sector: the extraordinary frequency becomes imaginary for
sufficiently transverse momenta. Additional causality requirements for
Pleba\'nski-type nonlinear electrodynamics may further restrict possible
phenomenological realizations
\cite{Schellstede2016,RussoTownsend2024,Abalos2015NLEDHyperbolic}.
Those restrictions are not the organizing criterion of the present
work; the result established here is that, in this class of
spontaneously Lorentz-breaking Hamiltonian NLED models, the Casimir
problem is intrinsically sector dependent.

\section{Conclusions}
\label{sec:conclusions}

We have studied the Casimir effect in a class of gauge-invariant
nonlinear electrodynamics models constructed to admit magnetic vacua
that spontaneously break Lorentz symmetry. The analysis was carried out
in the Pleba\'nski first-order Hamiltonian formulation, where the
nonlinear theory is specified by a single-invariant structural
potential \(\Vhat(P)\) and the vacuum problem is formulated in terms of
the reduced effective Hamiltonian.

The central object controlling the analysis is the magnetic-branch
function
\begin{equation}
\Sm(P)
=
\Vhat_P(P)+2P\Vhat_{PP}(P).
\end{equation}
Its vanishing at \(P=P_\star\) selects the Lorentz-breaking magnetic
vacuum of the Hamiltonian model and defines a rank-changing surface of
the associated constrained system. The same quantity also controls the
longitudinal magnetic response and the extraordinary fluctuation branch
in the regular sector. This identification is the main reason why the
Casimir problem considered here is not simply the standard Casimir
effect in a prescribed anisotropic medium.

We first quantized the electromagnetic fluctuations around a regular
purely magnetic background \(\bar P\), with \(\Sm(\bar P)\neq0\). In this
constant-rank sector the spectrum contains two propagating branches: an
ordinary Maxwell-like polarization,
\begin{equation}
\omega_1^2=k_\parallel^2+k_\perp^2,
\end{equation}
and an extraordinary anisotropic polarization,
\begin{equation}
\omega_2^2=k_\parallel^2+\alpha k_\perp^2,
\qquad
\alpha(\bar P)=\frac{\Vhat_P(\bar P)}{\Sm(\bar P)}.
\end{equation}
For the locally stable symmetry-breaking branches considered here,
\(\Vhat_P(P_\star)<0\), and the hyperbolic real-frequency side of the
regular sector adjacent to the degenerate surface corresponds to
\(\alpha(\bar P)\to+\infty\).

The regular-sector Casimir energy depends strongly on the orientation
of the magnetic background relative to the plates. When the regular
magnetic background is perpendicular to the plates, the anisotropy
rescales only the continuous momentum measure, giving
\begin{equation}
\frac{E_{\rm Cas}^{\perp}}{A}
=
-\frac{\pi^2}{1440a^3}
\left(
1+\frac{1}{\alpha}
\right).
\end{equation}
When the regular magnetic background is parallel to the plates, the
anisotropy also rescales the discrete cavity momentum, yielding
\begin{equation}
\frac{E_{\rm Cas}^{\parallel}}{A}
=
-\frac{\pi^2}{1440a^3}
(1+\alpha).
\end{equation}
Thus the parallel configuration develops a divergent regular-sector
contribution as \(\alpha\to+\infty\), whereas the perpendicular
configuration remains finite.

This divergence is not interpreted as the physical Casimir force of an
exact vacuum state. It is the singular endpoint of the regular
constant-rank sector of the chosen Hamiltonian NLED model. At
\(\Sm(P_\star)=0\), the longitudinal magnetic response loses rank and
the coupled sector must be analyzed after imposing the degenerate
condition. The exact degenerate fluctuation problem then contains only
the ordinary Maxwell-like propagating branch for generic momenta.

The analysis also shows that the formula obtained for \(\alpha>0\)
cannot be continued through the degenerate surface as another regular
zero-point spectrum. On the side where \(\alpha<0\), the extraordinary
dispersion relation develops imaginary frequencies for sufficiently
transverse momenta, signaling loss of hyperbolicity rather than a
regular Casimir regime. Together with the nonuniform behavior of the
\(\alpha\to+\infty\) limit, this indicates that the regular two-branch
optical sector has reached the boundary of its domain of applicability
within the model.

The apparent Casimir divergence is therefore a diagnostic of the
noncommutativity summarized in Eq.~\eqref{eq:noncommutativity}. It
signals a change in the number of propagating degrees of freedom at a
Hamiltonian-degenerate surface, rather than a conventional
vacuum-energy correction evaluated at the exact Lorentz-breaking state.
From this perspective, the Casimir effect becomes a way to test the
sector structure of nonlinear gauge theories constructed to realize
spontaneous Lorentz symmetry breaking. The main lesson is that, once
such an electrodynamics is accepted as the model under study, its
quantum-vacuum observables are sensitive not only to modified
dispersion relations, but also to rank-changing surfaces where the
regular optical description and the exact Hamiltonian fluctuation
problem cease to be equivalent.

\section*{Declarations}

\paragraph{Funding}
A.M.-R. acknowledges financial support from UNAM-PAPIIT project
No. IG100224, UNAM-PAPIME project No. PE109226, SECIHTI project
No. CBF-2025-I-1862, and the Marcos Moshinsky Foundation. R.L.
acknowledges partial support from CONAHCyT-Mexico under Grant
No. CBF-2023-2024-1937.

\paragraph{Competing interests}
The authors declare that they have no competing interests.

\paragraph{Data availability}
No datasets were generated or analyzed during the current study.

\appendix

\section{Cavity eigenmodes in the regular anisotropic sector}
\label{app:cavity_modes}

In this appendix we justify the standing-wave quantization used in the
parallel-plate calculation. The point is slightly more delicate than in
ordinary Maxwell theory because the perfect-conductor boundary
conditions are imposed on the physical fields $\delta\mathbf E$ and
$\delta\mathbf B$, whereas the Pleba\'nski formulation uses
$\delta\mathbf D$ and $\delta\mathbf H$ as Hamiltonian variables.

We work throughout this appendix in the regular propagation sector,
\begin{equation}
p\equiv \Vhat_P(\bar P)\neq0,
\qquad
s\equiv \Sm(\bar P)\neq0.
\label{eq:app_regular_coefficients}
\end{equation}
The linearized constitutive relations are
\begin{equation}
\delta\mathbf E=-p\,\delta\mathbf D,
\label{eq:app_E_constitutive}
\end{equation}
and
\begin{equation}
\delta\mathbf B
=
-p\,\delta\mathbf H_\perp
-
s\,\delta H_\parallel\,\hat{\mathbf n},
\label{eq:app_B_constitutive}
\end{equation}
where the perpendicular and longitudinal components are defined with
respect to the regular magnetic background direction
$\hat{\mathbf n}$.

The plates are located at $z=0$ and $z=a$, with normal
$\hat{\mathbf z}$. The perfect-conductor boundary conditions are
\begin{equation}
\hat{\mathbf z}\times\delta\mathbf E=0,
\qquad
\hat{\mathbf z}\cdot\delta\mathbf B=0,
\qquad
z=0,a.
\label{eq:app_pc_conditions}
\end{equation}
Using Eq.~\eqref{eq:app_E_constitutive}, the tangential electric
condition is equivalent, for $p\neq0$, to
\begin{equation}
\delta D_x=0,
\qquad
\delta D_y=0,
\qquad
z=0,a.
\label{eq:app_D_tangential}
\end{equation}

Let the magnetic fluctuation have harmonic time dependence and be
translationally invariant along the plates,
\begin{equation}
\delta\mathbf H(\mathbf x,t)
=
\mathbf h(z)\,
e^{i(k_xx+k_yy-\omega t)}.
\label{eq:app_H_profile}
\end{equation}
The linearized Ampère equation,
\begin{equation}
\nabla\times\delta\mathbf H=\partial_t\delta\mathbf D,
\label{eq:app_ampere}
\end{equation}
then gives, up to an irrelevant common factor,
\begin{equation}
\delta D_x
\propto
ik_y h_z-\frac{d h_y}{dz},
\qquad
\delta D_y
\propto
\frac{d h_x}{dz}-ik_x h_z.
\label{eq:app_D_from_H}
\end{equation}
Therefore Eq.~\eqref{eq:app_D_tangential} requires
\begin{equation}
ik_y h_z-\frac{d h_y}{dz}=0,
\qquad
\frac{d h_x}{dz}-ik_x h_z=0,
\qquad
z=0,a.
\label{eq:app_boundary_intermediate}
\end{equation}

We now specialize to the two orientations used in the main text.

\subsection{Regular magnetic background perpendicular to the plates}

For a regular magnetic background normal to the plates,
\begin{equation}
\hat{\mathbf n}=\hat{\mathbf z},
\label{eq:app_n_perp}
\end{equation}
the normal magnetic boundary condition follows from
\begin{equation}
\delta B_z=-s\,h_z.
\label{eq:app_Bz_perp}
\end{equation}
Since $s\neq0$ in the regular sector, $\delta B_z=0$ implies
\begin{equation}
h_z=0,
\qquad
z=0,a.
\label{eq:app_hz_zero_perp}
\end{equation}
Substituting this into Eq.~\eqref{eq:app_boundary_intermediate} gives
\begin{equation}
\frac{d h_x}{dz}=0,
\qquad
\frac{d h_y}{dz}=0,
\qquad
h_z=0,
\qquad
z=0,a.
\label{eq:app_bc_reduced_perp}
\end{equation}
The compatible standing-wave profiles may therefore be chosen as
\begin{equation}
h_x(z),h_y(z)\propto\cos(k_zz),
\qquad
h_z(z)\propto\sin(k_zz),
\label{eq:app_profiles_perp}
\end{equation}
with
\begin{equation}
k_z=\frac{n\pi}{a},
\qquad
n=0,1,2,\ldots.
\label{eq:app_kz_perp}
\end{equation}
The $n=0$ sector does not contribute to the separation-dependent part
of the regularized vacuum energy relevant for the force. It is
therefore omitted from the finite Casimir contribution, and one may
take $n=1,2,\ldots$ in the mode sums.

For this orientation,
\begin{equation}
k_\parallel=k_z,
\qquad
k_\perp^2=k_x^2+k_y^2.
\label{eq:app_k_components_perp}
\end{equation}
Substitution into the two regular bulk branches gives
\begin{equation}
\omega_{1,n}^{\perp}
=
\sqrt{
k_x^2+k_y^2+\left(\frac{n\pi}{a}\right)^2
},
\label{eq:app_omega1_perp}
\end{equation}
and
\begin{equation}
\omega_{2,n}^{\perp}
=
\sqrt{
\left(\frac{n\pi}{a}\right)^2
+
\alpha(k_x^2+k_y^2)
}.
\label{eq:app_omega2_perp}
\end{equation}
These are the frequencies used in the perpendicular configuration.

\subsection{Regular magnetic background parallel to the plates}

For a regular magnetic background parallel to the plates, chosen along
the $x$ axis,
\begin{equation}
\hat{\mathbf n}=\hat{\mathbf x},
\label{eq:app_n_parallel}
\end{equation}
the normal component of the magnetic induction is
\begin{equation}
\delta B_z=-p\,h_z.
\label{eq:app_Bz_parallel}
\end{equation}
Since $p\neq0$, the condition $\delta B_z=0$ again implies
\begin{equation}
h_z=0,
\qquad
z=0,a.
\label{eq:app_hz_zero_parallel}
\end{equation}
The tangential electric boundary conditions then reduce once more to
\begin{equation}
\frac{d h_x}{dz}=0,
\qquad
\frac{d h_y}{dz}=0,
\qquad
h_z=0,
\qquad
z=0,a.
\label{eq:app_bc_reduced_parallel}
\end{equation}
Thus the allowed normal momenta are again
\begin{equation}
k_z=\frac{n\pi}{a},
\qquad
n=0,1,2,\ldots.
\label{eq:app_kz_parallel}
\end{equation}
As in the perpendicular configuration, the $n=0$ contribution does not
affect the separation-dependent part of the regularized vacuum energy
and is excluded from the finite Casimir force.

For this orientation,
\begin{equation}
k_\parallel=k_x,
\qquad
k_\perp^2=k_y^2+k_z^2.
\label{eq:app_k_components_parallel}
\end{equation}
Therefore the two regular-sector frequencies become
\begin{equation}
\omega_{1,n}^{\parallel}
=
\sqrt{
k_x^2+k_y^2+\left(\frac{n\pi}{a}\right)^2
},
\label{eq:app_omega1_parallel}
\end{equation}
and
\begin{equation}
\omega_{2,n}^{\parallel}
=
\sqrt{
k_x^2+
\alpha\left[
k_y^2+\left(\frac{n\pi}{a}\right)^2
\right]
}.
\label{eq:app_omega2_parallel}
\end{equation}
These are the frequencies used in the parallel configuration.

The derivation above shows that, for the two special orientations
considered in the main text, the perfect-conductor boundary conditions
are compatible with the standard standing-wave quantization of the
normal momentum. The anisotropy modifies the dispersion relation, and
therefore the orientation dependence of the zero-point energy, but it
does not modify the allowed values of $k_z$ in the regular sector.

This conclusion relies essentially on $s=\Sm(\bar P)\neq0$. It should not
be extrapolated directly to the exact degenerate surface
$\Sm(P_\star)=0$. In particular, when the magnetic background is
perpendicular to the plates, the normal magnetic boundary condition
$\delta B_z=-s\,h_z=0$ no longer constrains $h_z$ once $s=0$. This is
another manifestation of the fact that the exact degenerate vacuum must
be quantized separately, rather than obtained by a naive continuation
of the regular cavity spectrum.

\bibliographystyle{spphys}
\bibliography{biblio2}

@article{BluhmGagnePottingVrublevskis2008,
  author        = {Bluhm, Robert and Gagne, Nolan L. and Potting, Robertus and Vrublevskis, Arturs},
  title         = {Constraints and Stability in Vector Theories with Spontaneous Lorentz Violation},
  journal       = {Phys. Rev. D},
  volume        = {77},
  pages         = {125007},
  year          = {2008},
  doi           = {10.1103/PhysRevD.77.125007},
  eprint        = {0802.4071},
  archivePrefix = {arXiv},
  primaryClass  = {hep-th}
}

@article{BonderEscobar2016,
  author        = {Bonder, Yuri and Escobar, Carlos A.},
  title         = {Dynamical Ambiguities in Models with Spontaneous Lorentz Violation},
  journal       = {Phys. Rev. D},
  volume        = {93},
  pages         = {025020},
  year          = {2016},
  doi           = {10.1103/PhysRevD.93.025020},
  eprint        = {1510.05999},
  archivePrefix = {arXiv},
  primaryClass  = {hep-th}
}

@article{EscobarLinares2022,
  author        = {Escobar, C. A. and Linares, Rom{\'a}n},
  title         = {Spontaneous Symmetry Breaking in Models with Second-Class Constraints},
  journal       = {Phys. Rev. D},
  volume        = {106},
  pages         = {036027},
  year          = {2022},
  doi           = {10.1103/PhysRevD.106.036027},
  eprint        = {2207.05251},
  archivePrefix = {arXiv},
  primaryClass  = {hep-th}
}

@book{HenneauxTeitelboim1992,
  author    = {Henneaux, Marc and Teitelboim, Claudio},
  title     = {Quantization of Gauge Systems},
  publisher = {Princeton University Press},
  address   = {Princeton},
  year      = {1992}
}

@article{MiskovicZanelli2003,
  author        = {Miskovic, Olivera and Zanelli, Jorge},
  title         = {Dynamical Structure of Irregular Constrained Systems},
  journal       = {J. Math. Phys.},
  volume        = {44},
  pages         = {3876--3887},
  year          = {2003},
  doi           = {10.1063/1.1601299},
  eprint        = {hep-th/0302033},
  archivePrefix = {arXiv},
  primaryClass  = {hep-th}
}

@article{SaavedraTroncosoZanelli2001,
  author        = {Saavedra, J. and Troncoso, R. and Zanelli, J.},
  title         = {Degenerate Dynamical Systems},
  journal       = {J. Math. Phys.},
  volume        = {42},
  pages         = {4383--4390},
  year          = {2001},
  doi           = {10.1063/1.1389088},
  eprint        = {hep-th/0011231},
  archivePrefix = {arXiv},
  primaryClass  = {hep-th}
}

@article{MiskovicTroncosoZanelli2005,
  author        = {Miskovic, Olivera and Troncoso, Ricardo and Zanelli, Jorge},
  title         = {Canonical Sectors of Five-Dimensional Chern-Simons Theories},
  journal       = {Phys. Lett. B},
  volume        = {615},
  pages         = {277--284},
  year          = {2005},
  doi           = {10.1016/j.physletb.2005.04.043},
  eprint        = {hep-th/0504055},
  archivePrefix = {arXiv},
  primaryClass  = {hep-th}
}

@article{DeMicheliZanelli2012,
  author        = {de Micheli, Fiorenza and Zanelli, Jorge},
  title         = {Quantum Degenerate Systems},
  journal       = {J. Math. Phys.},
  volume        = {53},
  pages         = {102112},
  year          = {2012},
  doi           = {10.1063/1.4753996},
  eprint        = {1203.0022},
  archivePrefix = {arXiv},
  primaryClass  = {hep-th}
}

@article{DeserErtlGrumiller2013,
  author        = {Deser, S. and Ertl, S. and Grumiller, D.},
  title         = {Canonical Bifurcation in Higher Derivative, Higher Spin, Theories},
  journal       = {J. Phys. A},
  volume        = {46},
  pages         = {214018},
  year          = {2013},
  doi           = {10.1088/1751-8113/46/21/214018},
  eprint        = {1208.0339},
  archivePrefix = {arXiv},
  primaryClass  = {hep-th}
}

@article{ArkaniHamed2004GhostCondensate,
  author        = {Arkani-Hamed, Nima and Cheng, Hsin-Chia and Luty, Markus A. and Mukohyama, Shinji},
  title         = {Ghost condensation and a consistent infrared modification of gravity},
  journal       = {J. High Energy Phys.},
  volume        = {2004},
  number        = {05},
  pages         = {074},
  year          = {2004},
  doi           = {10.1088/1126-6708/2004/05/074},
  eprint        = {hep-th/0312099},
  archivePrefix = {arXiv},
  primaryClass  = {hep-th}
}

@article{Hornreich1975Lifshitz,
  author  = {Hornreich, R. M. and Luban, M. and Shtrikman, S.},
  title   = {Critical Behavior at the Onset of k-Space Instability on the Lambda Line},
  journal = {Phys. Rev. Lett.},
  volume  = {35},
  pages   = {1678--1681},
  year    = {1975},
  doi     = {10.1103/PhysRevLett.35.1678}
}

@article{DeserWaldron2001PartialMasslessness,
  author        = {Deser, S. and Waldron, A.},
  title         = {Partial Masslessness of Higher Spins in (A)dS},
  journal       = {Nucl. Phys. B},
  volume        = {607},
  pages         = {577--604},
  year          = {2001},
  doi           = {10.1016/S0550-3213(01)00212-7},
  eprint        = {hep-th/0103198},
  archivePrefix = {arXiv},
  primaryClass  = {hep-th}
}

@article{Seifert2019SingularHamiltonians,
  author        = {Seifert, Michael D.},
  title         = {Singular Hamiltonians in Models with Spontaneous Lorentz Symmetry Breaking},
  journal       = {Phys. Rev. D},
  volume        = {100},
  pages         = {065017},
  year          = {2019},
  doi           = {10.1103/PhysRevD.100.065017},
  eprint        = {1903.06140},
  archivePrefix = {arXiv},
  primaryClass  = {hep-th}
}

@article{Abalos2015NLEDHyperbolic,
  author        = {Abalos, Fernando and Carrasco, Federico and Goulart, Erico and Reula, Oscar},
  title         = {Nonlinear Electrodynamics as a Symmetric Hyperbolic System},
  journal       = {Phys. Rev. D},
  volume        = {92},
  pages         = {084024},
  year          = {2015},
  doi           = {10.1103/PhysRevD.92.084024},
  eprint        = {1507.02262},
  archivePrefix = {arXiv},
  primaryClass  = {gr-qc}
}

@article{RussoTownsend2023,
  author        = {Russo, Jorge G. and Townsend, Paul K.},
  title         = {Nonlinear electrodynamics without birefringence},
  journal       = {JHEP},
  volume        = {01},
  pages         = {039},
  year          = {2023},
  doi           = {10.1007/JHEP01(2023)039},
  eprint        = {2211.10689},
  archivePrefix = {arXiv},
  primaryClass  = {hep-th}
}

@article{BialynickaBirula1970,
  author  = {Bialynicka-Birula, Zofia and Bialynicki-Birula, Iwo},
  title   = {Nonlinear effects in quantum electrodynamics. Photon propagation and photon splitting in an external field},
  journal = {Phys. Rev. D},
  volume  = {2},
  pages   = {2341--2345},
  year    = {1970},
  doi     = {10.1103/PhysRevD.2.2341}
}

@article{BornInfeld1934,
  author  = {Born, Max and Infeld, Leopold},
  title   = {Foundations of the new field theory},
  journal = {Proc. R. Soc. Lond. A},
  volume  = {144},
  number  = {852},
  pages   = {425--451},
  year    = {1934},
  doi     = {10.1098/rspa.1934.0059}
}

@article{ColladayKostelecky1997,
  author        = {Colladay, Don and Kosteleck{\'y}, V. Alan},
  title         = {{CPT} violation and the standard model},
  journal       = {Phys. Rev. D},
  volume        = {55},
  pages         = {6760--6774},
  year          = {1997},
  doi           = {10.1103/PhysRevD.55.6760},
  eprint        = {hep-ph/9703464},
  archivePrefix = {arXiv}
}

@article{ColladayKostelecky1998,
  author        = {Colladay, Don and Kosteleck{\'y}, V. Alan},
  title         = {Lorentz-violating extension of the standard model},
  journal       = {Phys. Rev. D},
  volume        = {58},
  pages         = {116002},
  year          = {1998},
  doi           = {10.1103/PhysRevD.58.116002},
  eprint        = {hep-ph/9809521},
  archivePrefix = {arXiv}
}

@article{Schellstede2016,
  author        = {Schellstede, Gerold O. and Perlick, Volker and L{\"a}mmerzahl, Claus},
  title         = {On causality in nonlinear vacuum electrodynamics of the Pleba{\'n}ski class},
  journal       = {Annalen der Physik},
  volume        = {528},
  number        = {9--10},
  pages         = {738--749},
  year          = {2016},
  doi           = {10.1002/andp.201600124},
  eprint        = {1604.02545},
  archivePrefix = {arXiv},
  primaryClass  = {gr-qc}
}

@article{RussoTownsend2024,
  author        = {Russo, Jorge G. and Townsend, Paul K.},
  title         = {Causality and energy conditions in nonlinear electrodynamics},
  journal       = {JHEP},
  volume        = {06},
  pages         = {191},
  year          = {2024},
  doi           = {10.1007/JHEP06(2024)191},
  eprint        = {2404.09994},
  archivePrefix = {arXiv},
  primaryClass  = {hep-th}
}

@article{CruzBezerraPetrov2017,
  author        = {Cruz, M. B. and Bezerra de Mello, E. R. and Petrov, A. Yu.},
  title         = {Casimir effects in Lorentz-violating scalar field theory},
  journal       = {Phys. Rev. D},
  volume        = {96},
  pages         = {045019},
  year          = {2017},
  doi           = {10.1103/PhysRevD.96.045019},
  eprint        = {1705.03331},
  archivePrefix = {arXiv},
  primaryClass  = {hep-th}
}

@article{CruzBezerraPetrov2018,
  author        = {Cruz, M. B. and Bezerra de Mello, E. R. and Petrov, A. Yu.},
  title         = {Thermal corrections to the Casimir energy in a Lorentz-breaking scalar field theory},
  journal       = {Mod. Phys. Lett. A},
  volume        = {33},
  number        = {20},
  pages         = {1850115},
  year          = {2018},
  doi           = {10.1142/S0217732318501158},
  eprint        = {1803.07446},
  archivePrefix = {arXiv},
  primaryClass  = {hep-th}
}

@article{CruzBezerraMota2020,
  author        = {Cruz, M. B. and Bezerra de Mello, E. R. and Santana Mota, H. F.},
  title         = {Casimir energy and topological mass for a massive scalar field with Lorentz violation},
  journal       = {Phys. Rev. D},
  volume        = {102},
  pages         = {045006},
  year          = {2020},
  doi           = {10.1103/PhysRevD.102.045006},
  eprint        = {2005.09513},
  archivePrefix = {arXiv},
  primaryClass  = {hep-th}
}

@article{BezerraCruz2023,
  author        = {Bezerra de Mello, E. R. and Cruz, M. B.},
  title         = {Scalar Casimir effects in a Lorentz violation scenario induced by the presence of constant vectors},
  journal       = {Int. J. Mod. Phys. A},
  volume        = {38},
  pages         = {2350062},
  year          = {2023},
  doi           = {10.1142/S0217751X23500628},
  eprint        = {2210.09243},
  archivePrefix = {arXiv},
  primaryClass  = {hep-th}
}

@article{EscobarMedelMartinRuiz2020,
  author        = {Escobar, C. A. and Medel, Leonardo and Mart{\'i}n-Ruiz, A.},
  title         = {Casimir effect in Lorentz-violating scalar field theory: A local approach},
  journal       = {Phys. Rev. D},
  volume        = {101},
  pages         = {095011},
  year          = {2020},
  doi           = {10.1103/PhysRevD.101.095011},
  eprint        = {2005.00151},
  archivePrefix = {arXiv},
  primaryClass  = {hep-th}
}

@article{EscobarMartinRuizFrancaGarcia2020,
  author        = {Escobar, C. A. and Mart{\'i}n-Ruiz, A. and Franca, O. J. and Garcia, Marcos A. G.},
  title         = {A non-perturbative approach to the scalar Casimir effect with Lorentz symmetry violation},
  journal       = {Phys. Lett. B},
  volume        = {807},
  pages         = {135567},
  year          = {2020},
  doi           = {10.1016/j.physletb.2020.135567},
  eprint        = {2005.14217},
  archivePrefix = {arXiv},
  primaryClass  = {hep-th}
}

@article{EscobarRuizCylinder2021,
  author        = {Escobar-Ruiz, A. M. and Mart{\'i}n-Ruiz, A. and Escobar, C. A. and Linares, Rom{\'a}n},
  title         = {Scalar Casimir effect for a conducting cylinder in a Lorentz-violating background},
  journal       = {Int. J. Mod. Phys. A},
  volume        = {36},
  pages         = {2150168},
  year          = {2021},
  doi           = {10.1142/S0217751X21501682},
  eprint        = {2105.12953},
  archivePrefix = {arXiv},
  primaryClass  = {hep-th}
}

@article{MartinRuizSphere2020,
  author        = {Mart{\'i}n-Ruiz, A. and Escobar, C. A. and Escobar-Ruiz, A. M. and Franca, O. J.},
  title         = {Lorentz violating scalar Casimir effect for a {$D$}-dimensional sphere},
  journal       = {Phys. Rev. D},
  volume        = {102},
  pages         = {015027},
  year          = {2020},
  doi           = {10.1103/PhysRevD.102.015027},
  eprint        = {2006.00696},
  archivePrefix = {arXiv},
  primaryClass  = {hep-th}
}

@book{Plebanski1970,
  author    = {Pleba{\'n}ski, J.},
  title     = {Lectures on Non-linear Electrodynamics},
  publisher = {NORDITA},
  address   = {Copenhagen},
  year      = {1970}
}

@article{EscobarPotting2020IJMPA,
  author  = {Escobar, C. A. and Potting, R.},
  title   = {Nonlinear vacuum electrodynamics and spontaneous breaking of Lorentz symmetry},
  journal = {Int. J. Mod. Phys. A},
  volume  = {35},
  pages   = {2050174},
  year    = {2020},
  doi     = {10.1142/S0217751X20501742},
  eprint  = {1810.01677},
  archivePrefix = {arXiv},
  primaryClass = {hep-th}
}

@article{HeisenbergEuler1936,
  author  = {Heisenberg, W. and Euler, H.},
  title   = {Folgerungen aus der Diracschen Theorie des Positrons},
  journal = {Z. Phys.},
  volume  = {98},
  pages   = {714--732},
  year    = {1936},
  doi     = {10.1007/BF01343663}
}

@article{Schwinger1951,
  author  = {Schwinger, J.},
  title   = {On Gauge Invariance and Vacuum Polarization},
  journal = {Phys. Rev.},
  volume  = {82},
  pages   = {664--679},
  year    = {1951},
  doi     = {10.1103/PhysRev.82.664}
}

@article{Bandos2020,
  author  = {Bandos, Igor and Lechner, Kurt and Sorokin, Dmitri and Townsend, Paul K.},
  title   = {A non-linear duality-invariant conformal extension of Maxwell's equations},
  journal = {Phys. Rev. D},
  volume  = {102},
  pages   = {121703},
  year    = {2020},
  doi     = {10.1103/PhysRevD.102.121703},
  eprint  = {2007.09092},
  archivePrefix = {arXiv},
  primaryClass = {hep-th}
}

@article{Sorge2024,
  author  = {Sorge, Francesco},
  title   = {Casimir effect in ModMax and Euler-Heisenberg electrodynamics},
  journal = {Phys. Rev. D},
  volume  = {110},
  pages   = {116006},
  year    = {2024},
  doi     = {10.1103/PhysRevD.110.116006}
}

@article{PlacidoFlores2026Stable,
  author        = {Pl{\'a}cido-Flores, E. and Linares, R. and L{\'o}pez, V. and Escobar, C. A.},
  title         = {Stable magnetic Lorentz-violating vacua in gauge-invariant nonlinear electrodynamics},
  journal       = {Eur. Phys. J. C},
  volume        = {86},
  pages         = {619},
  year          = {2026},
  doi           = {10.1140/epjc/s10052-026-15892-w},
  eprint        = {2605.03341},
  archivePrefix = {arXiv},
  primaryClass  = {hep-th}
}

@misc{EscobarLinares2026Constitutive,
  author        = {C. A. Escobar and Rom{\'a}n Linares},
  title         = {Constitutive Origin of Hamiltonian Degeneracy in Nonlinear Electrodynamics with Spontaneous Lorentz Symmetry Breaking},
  eprint        = {2605.14172},
  archivePrefix = {arXiv},
  primaryClass  = {hep-th},
  year          = {2026},
  note          = {arXiv:2605.14172 [hep-th]}
}

@article{Casimir1948,
  author = {Casimir, H. B. G.},
  title = {On the Attraction Between Two Perfectly Conducting Plates},
  journal = {Proceedings of the Koninklijke Nederlandse Akademie van Wetenschappen},
  volume = {51},
  pages = {793--795},
  year = {1948}
}

@book{Milton2001,
  author    = {Milton, K. A.},
  title     = {The Casimir Effect},
  publisher = {World Scientific},
  address   = {Singapore},
  year      = {2001}
}

@article{Boillat1970,
  author = {Boillat, G.},
  title = {Nonlinear Electrodynamics: Lagrangians and Equations of Motion},
  journal = {Journal of Mathematical Physics},
  volume = {11},
  number = {3},
  pages = {941--951},
  year = {1970},
  doi = {10.1063/1.1665231}
}

@article{Novello2000,
  author = {Novello, M. and De Lorenci, V. A. and Salim, J. M. and Klippert, R.},
  title = {Geometrical Aspects of Light Propagation in Nonlinear Electrodynamics},
  journal = {Physical Review D},
  volume = {61},
  pages = {045001},
  year = {2000},
  doi = {10.1103/PhysRevD.61.045001}
}

@article{ObukhovRubilar2002,
  author = {Obukhov, Y. N. and Rubilar, G. F.},
  title = {Fresnel Analysis of Wave Propagation in Nonlinear Electrodynamics},
  journal = {Physical Review D},
  volume = {66},
  pages = {024042},
  year = {2002},
  doi = {10.1103/PhysRevD.66.024042}
}
\end{document}